# Development of a strontium optical lattice clock for the SOC mission on the ISS

S. Origlia*[a], S. Schiller[a], M.S. Pramod[a], L. Smith[b], Y. Singh[b], W. He[b], S. Viswam[b], D. Świerad[b], J. Hughes[b], K. Bongs[b], U. Sterr[c], Ch. Lisdat[c], S. Vogt[c], S. Bize[d], J. Lodewyck[d], R. Le Targat[d], D. Holleville[d], B. Venon[d], P. Gill[e], G. Barwood[e], I. R. Hill[e], Y. Ovchinnikov[e], A. Kulosa[f], W. Ertmer[f], E.-M. Rasel[f], J. Stuhler[g], W. Kaenders[g], *and the SOC2 consortium contributors*

[a]Institut für Experimentalphysik, Heinrich-Heine-Universität Düsseldorf (HHU), D 40225 Düsseldorf;
[b]School of Physics & Astronomy, University of Birmingham (UoB), UK B15 2TT Birmingham;
[c]Physikalisch-Technische Bundesanstalt (PTB), Bundesallee 100, D 38116 Braunschweig;
[d]SYRTE, Observatoire de Paris, Avenue de l'Observatoire 61, F 75014 Paris;
[e]National Physical Laboratory (NPL), Hampton Road, UK TW11 0LW Teddington;
[f]Institut für Quantenoptik, Leibniz Universität Hannover (LUH), D 30167 Hannover;
[g]TOPTICA Photonics AG, Lochhamer Schlag 19, D 82166 Gräfelfing

## ABSTRACT

The ESA mission "Space Optical Clock" project aims at operating an optical lattice clock on the ISS in approximately 2023. The scientific goals of the mission are to perform tests of fundamental physics, to enable space-assisted relativistic geodesy and to intercompare optical clocks on the ground using microwave and optical links. The performance goal of the space clock is less than $1 \times 10^{-17}$ uncertainty and $1 \times 10^{-15} \tau^{-1/2}$ instability. Within an EU-FP7-funded project, a strontium optical lattice clock demonstrator has been developed. Goal performances are instability below $1 \times 10^{-15} \tau^{-1/2}$ and fractional inaccuracy $5 \times 10^{-17}$. For the design of the clock, techniques and approaches suitable for later space application are used, such as modular design, diode lasers, low power consumption subunits, and compact dimensions. The Sr clock apparatus is fully operational, and the clock transition in $^{88}$Sr was observed with linewidth as small as 9 Hz.

## 1. INTRODUCTION

The development of optical clocks worldwide is opening the door to new applications, not only with clocks on ground but also in space. The most recent results in the field of optical clocks show that uncertainty and instability approaching $1 \times 10^{-18}$ has been achieved in neutral strontium[1] and ytterbium[2] clocks. One current focus is transportable optical clocks. They are desirable for several reasons, such as comparisons between frequency standards in distant laboratories and for relativistic geodesy, and they also represent an essential step towards space optical clocks.

Within the Space Optical Clock[3] (SOC) mission development two transportable optical lattice clocks have been built, based on neutral ytterbium[4] and strontium atoms. The first strontium breadboard for the SOC project was developed starting in 2007 in Firenze[5] and from 2013 on a more advanced setup has been assembled in Birmingham[6]. The latter's Sr atomic package and corresponding laser ensemble were moved to PTB (Braunschweig) in mid-2015 for integration with the custom-made clock laser, and for metrological characterization. The system is currently fully operational with $^{88}$Sr, showing a clock transition spectral linewidth of the order of 40 Hz using the custom-made clock laser. An even narrower linewidth (9 Hz) was achieved by locking the clock laser to a non-transportable ultrastable cavity.

*stefano.origlia@hhu.de; website: www.soc2.eu

This first section of this report will introduce the applications of atomic clocks in space and the Space Optical Clock project, as well as the basics of a strontium optical lattice clock. The second section is focused on the design of the SOC strontium clock. In the third section we report the first results. Finally, the fourth section will outline the future developments.

**1.1 Space clocks applications**

Optical clocks in space have a wide range of technological and scientific applications. First of all in metrology: a space-based clock, together with high performance link, can be used as "master clock" to compare the highest-performance clocks on Earth with a relative frequency uncertainty at $10^{-18}$ level, to create an international optical atomic time scale towards a future redefinition of the SI second. A space clock will also help to explore the limits of our fundamental physical laws, with experiments at the interface between general relativity and quantum mechanics. The gravitational redshift in the Earth field can be measured by comparison between the space-based clock and ground clocks, while comparing terrestrial clocks at large distance in east-west direction via the space-based clock will allow to measure the redshift in the gravitational field of the Sun[7,8,9,10,11,12]. A precision mapping of Earth's gravitational potential will be possible by relativistic geodesy[13]. It is based on a comparison between the frequencies of the space clock and ground-based clocks located at geophysically interesting sites, combined with an accurate determination of the orbit of the space clock. It is expected that the gravitational potential at any location on Earth can be measured with resolution equivalent to a 1 cm height difference within a 1-day long measurement time. Finally, a space-based atomic clock might also be useful for supporting for deep-space navigation.

**1.2 The Space Optical Clock project**

"Space Optical Clock" is an ESA mission since 2006. It aims to operate an optical clock on the International Space Station (ISS) by 2023. This mission is the natural follow-on of the mission ACES[14] (Atomic Clock Ensemble in Space, launch 2017), that will operate a cold-atom microwave clock on the ISS. Compared with ACES, the SOC mission will have performance improved by a factor of at least 10 in both clock and frequency link. The relative inaccuracy will be at $1 \times 10^{-17}$ level; the instability $1 \times 10^{-15}$ at one second interrogation time.

The SOC project (2007 – 2010), within the ELIPS-3 program of ESA, and the following SOC2 project[15] (2011 – 2015; EU-FP7-SPACE-2010-1 project no. 263500), were the first experimental studies of the feasibility of an optical lattice clock for space applications. Within the SOC projects two transportable optical lattice clocks based on neutral ytterbium and strontium atoms were developed.

The main requirements for the two clocks are: performance, compact size and moderate mass, robustness and low power consumption. To match these requirements, novel solutions have been used for the implementation of the clocks, such modular laser systems, containing compact and transportable laser sources[6,15,16], use of a low-power oven as atomic source[17], a permanent-magnet Zeeman slower[18], a compact and lightweight vacuum chamber[6], a robust laser frequency stabilization system[19] and a robust clock laser reference cavity[6].

**1.3 Strontium lattice clocks**

Strontium is an alkaline-earth-metal atom. As other elements in the same group, such as Mg and Ca, it has two electrons in the outer shell and an electronic structure composed of singlet and triplet states (Figure 1).

The strong, dipole-allowed $^1S_0$-$^1P_1$ transition at 461 nm, with a linewidth of 32 MHz, is used to slow down the atoms in the thermal atomic beam produced by the oven, and to trap them into a magneto-optical trap (MOT). Since the temperature of the atoms in this first-stage MOT (2 – 3 mK) is not low enough for high precision spectroscopy, a second-stage MOT is necessary, operating on the $^1S_0 - {}^3P_1$ transition (wavelength 689 nm). After this stage, the temperature of the atoms is at 1 µK level. In contrast to $^{88}$Sr, for an efficient cooling of $^{87}$Sr in the second-stage MOT, a second ("stirring") laser emitting at a frequency shifted by a few GHz with respect of the cooling laser is required[20] (see Section 2.1).

After cooling down the atoms to µK temperature, at which the atoms' velocity is still on the order of a few cm/s, they can be trapped in an optical lattice. This results in sub-µm spatial confinement of the atoms in one spatial direction, then allowing spectroscopy of ultra-narrow resonance lines, free of 1st-order Doppler effect. The wavelength of the light used for the realization of the lattice is properly selected to minimize the light shift on the clock transition due to the perturbation induced by the lattice's electric field on the energy of the clock states. For Sr this wavelength is $\lambda_L$=813 nm.

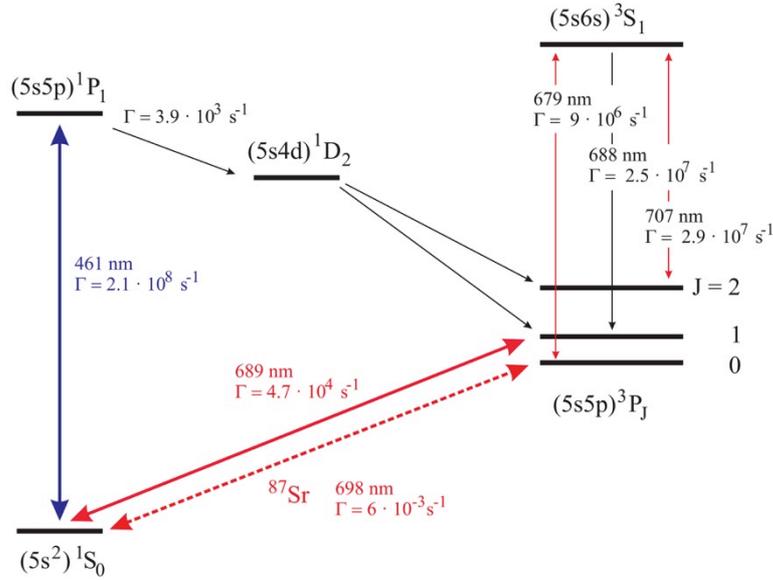

Figure 1. Simplified energy diagram for strontium, including relevant optical transitions and spontaneous decay rates.

The clock transition is the ultra-narrow $^1S_0 - {}^3P_0$ transition at 698 nm. It is interrogated by a clock laser whose spectral linewidth is narrowed down to below 1 Hz by frequency-locking to an ultra-stable reference cavity (see Section 2.3).

Finally, two more lasers at 679 nm and 707 nm are used to improve the efficiency of the first-stage MOT. These lasers repump the atoms from the states $^3P_0$ and $^3P_2$, which have a long lifetime, to the ground state through the $^3S_1$ level.

The reason why strontium was chosen as good candidate for space application is that all the wavelengths of interest are accessible by available semiconductor lasers and extensive experience with Sr optical clocks is available in European metrology laboratories.

Strontium has two isotopes used for optical clocks, a bosonic one ($^{88}$Sr) and a fermionic one ($^{87}$Sr). Since the bosonic isotope has zero nuclear spin, it is less sensitive to stray magnetic fields; furthermore its isotopic abundance (82.6%) is much higher than that of $^{87}$Sr (7.0%), and thus cooling and trapping of $^{88}$Sr is easier to achieve. On the other hand, in $^{87}$Sr s-wave scattering (a particular type of atom-atom collision) in a 1D lattice is suppressed, while for $^{88}$Sr the atoms have to be confined properly (i.e. using a 3D lattice) to remove the effect of collisions.

## 2. THE SOC2 STRONTIUM BREADBOARD

Figure 2 gives an overview of the clock apparatus including the atomic unit, which consists of the cooling laser system, the breadboard with the vacuum chamber, and the clock laser. The atomic unit is placed in a rack with a total volume (excluding electronics) of 970 liter. In this section each of these units will be briefly described.

### 2.1 Laser systems

The modular laser system (Figure 3) is conceived for easy intervention on subsystems and an easy implementation of new laser sources, for example space prototypes developed by industries within the technology development phase of the space clock. In fact, all lasers are interconnected and delivered to the "science" chamber by optical fibers. The Frequency Stabilization System[19] (FSS) is used to stabilize the 461, 689 and 813 nm lasers (see below). All the laser heads, apart for the one of the clock laser, are commercial diode lasers. The laser breadboards contain half-inch optics components, including ultra-stable mirror mounts.

The laser for the first-stage MOT is a TOPTICA SHG-TA-pro that provides up to 500 mW through frequency doubling of 922 nm. The 922 nm output is stabilized to the FSS. The fiber output of the laser is connected to a compact frequency distribution unit (size 30 cm × 20 cm × 10 cm, mass 5 kg) used to generate the three MOT beams, the slower beam (used to slow down the thermal atoms from the oven before they reach the capture region in the MOT) and the detection beam, for the fluorescence spectroscopy of the trapped atoms.

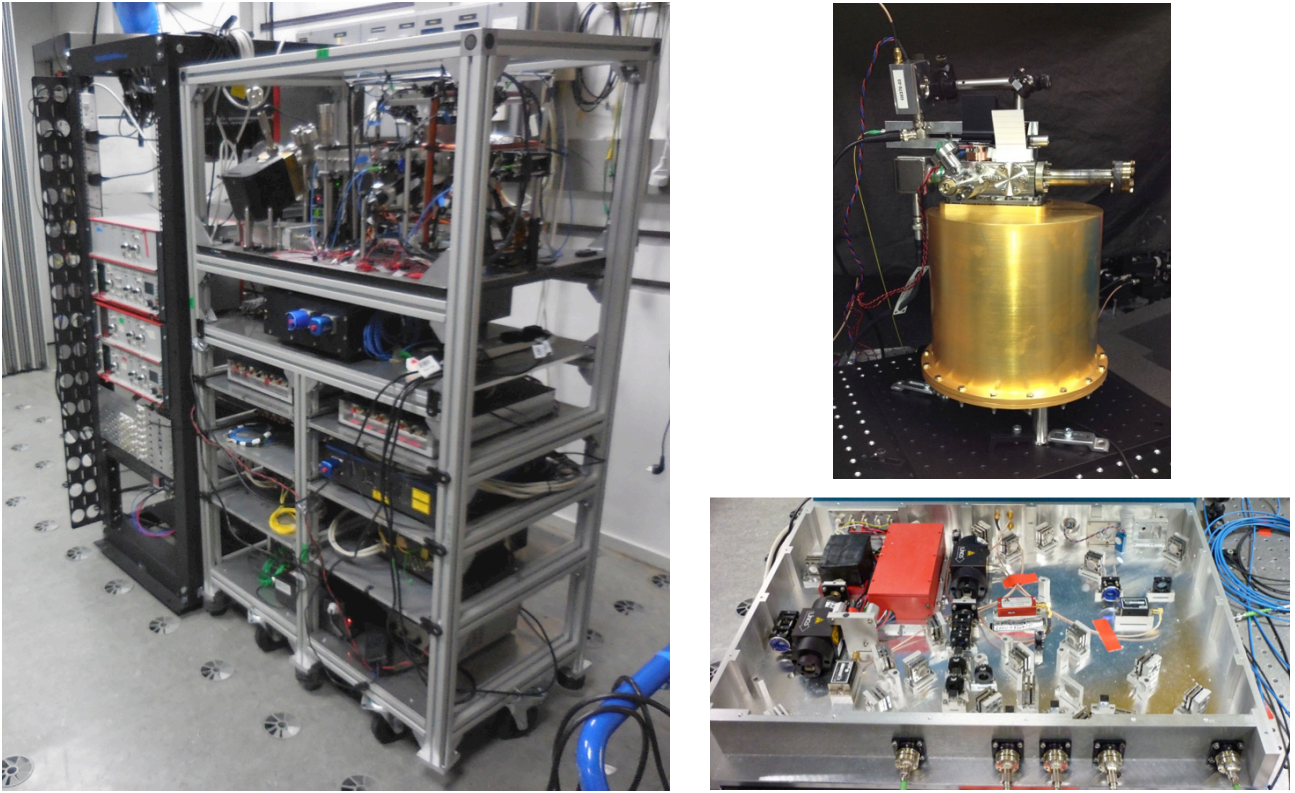

Figure 2. Left: the Sr atomics unit including laser systems, atomics package (top breadboard) and electronics. Right: clock laser cavity (top) and clock laser breadboard (bottom). To view the cavity in operation while it is rotated by hand by 180 degree, see the video clip[21].

For the second-stage cooling, the natural linewidth of the $^1S_0 - {}^3P_1$ transition of 7.5 kHz necessitates a 689 nm laser with approximately 1 kHz linewidth. To achieve this, the laser is frequency-locked to the FSS. To cover a wide range of atomic velocities when the atoms are transferred from the first-stage to the second-stage MOT, the laser spectrum can be broadened with amplitude of up to 5 MHz (at 30 kHz modulation frequency). However, since the intensity of the MOT beams, more than 5 mW/cm$^2$, is high compared to the saturation intensity of the $^1S_0 - {}^3P_1$ transition (3.0 μW/cm$^2$), at the moment there is no need for such a broadening. The power broadening is sufficient to efficiently transfer the atoms from the first-stage to the second-stage MOT. The stabilized laser light from the 689 nm diode laser is used to inject a slave diode laser that can provide more than 30 mW. As mentioned above, a second laser at 689 nm (stirring laser) is required. The stirring laser is phase-locked to the cooling laser using a beat note generated between the two at 1.4 GHz. For these two lasers, commercial diode lasers (TOPTICA DL-pro) with long extended cavities, minimizing the free-running linewidth, are directly integrated (without housing) in compact breadboards (size 30 cm × 45 cm × 12 cm, mass 12 kg) with all the optics and components necessary for the injection-locking (cooling laser) and offset-locking (stirring laser), as well as for the control of the output beams (AOMs and shutters).

For the 1D optical dipole trap we use a master-oscillator-power-amplifier laser (TOPTICA TA-pro) that can provide up to 2 W at 813 nm. The laser is implemented in a 50 cm × 30 cm × 20 cm unit including a breadboard to generate two outputs, one for the locking to the FSS and one that goes to the MOT chamber. To filter out the spurious spectral background generated by the amplified spontaneous emission, a grating is placed at the output of the laser.

The two repumpers (wavelength 679 nm and 707 nm) are TOPTICA DL-pro diode lasers implemented in two identical units (size 26.5 cm × 32 cm × 12.1 cm, mass 15 kg). Each unit has two outputs, one to the vacuum chamber and one to a wavemeter for monitoring, or, if needed (i.e. if the passive stability of the lasers is not sufficient), for locking.

The 461 nm, 689 nm and 813 nm lasers are stabilized using the Frequency Stabilization System[19] (FSS). The FSS contains a monolithic ultra-low expansion (ULE) block consisting of three 10 cm-long cavities. The first cavity is for the 922 nm and 813 nm lasers: the measured linewidth for these two lasers after locking is less than 1 MHz. On the second cavity the 689 nm laser is stabilized, with a linewidth of less than 1 kHz (1 min time scale) and a drift of the order of 0.5 Hz/s (i.e. 10 – 20 kHz daily drift). Since this drift is larger than the transition's natural linewidth (7 kHz), the laser frequency can be corrected with a frequency offset obtained by interrogating the second cavity also by the independently stabilized clock laser (698 nm). The third cavity can be used for locking the 679 nm and 707 nm lasers, but is not used at the moment since the short-term passive frequency stability of these lasers is good enough. The total mass of the FSS is 25 kg, with a volume of 50 cm × 30 cm × 20 cm, including the vacuum chamber for the cavity and the waveguide phase modulators placed in a box. The lasers are sent into the FSS vacuum chamber via fiber feed-throughs. Since the cavities are not tunable, an offset-locking technique is used: the light from the respective laser is phase-modulated, generating two sidebands. One of the two sidebands is locked to the reference cavity by acting on the laser current. By tuning the DDS frequency producing the phase-modulation, one can tune the laser carrier frequency without loss of lock. This method can for example be used for frequency broadening the 689 nm light for the initial second-stage MOT cooling.

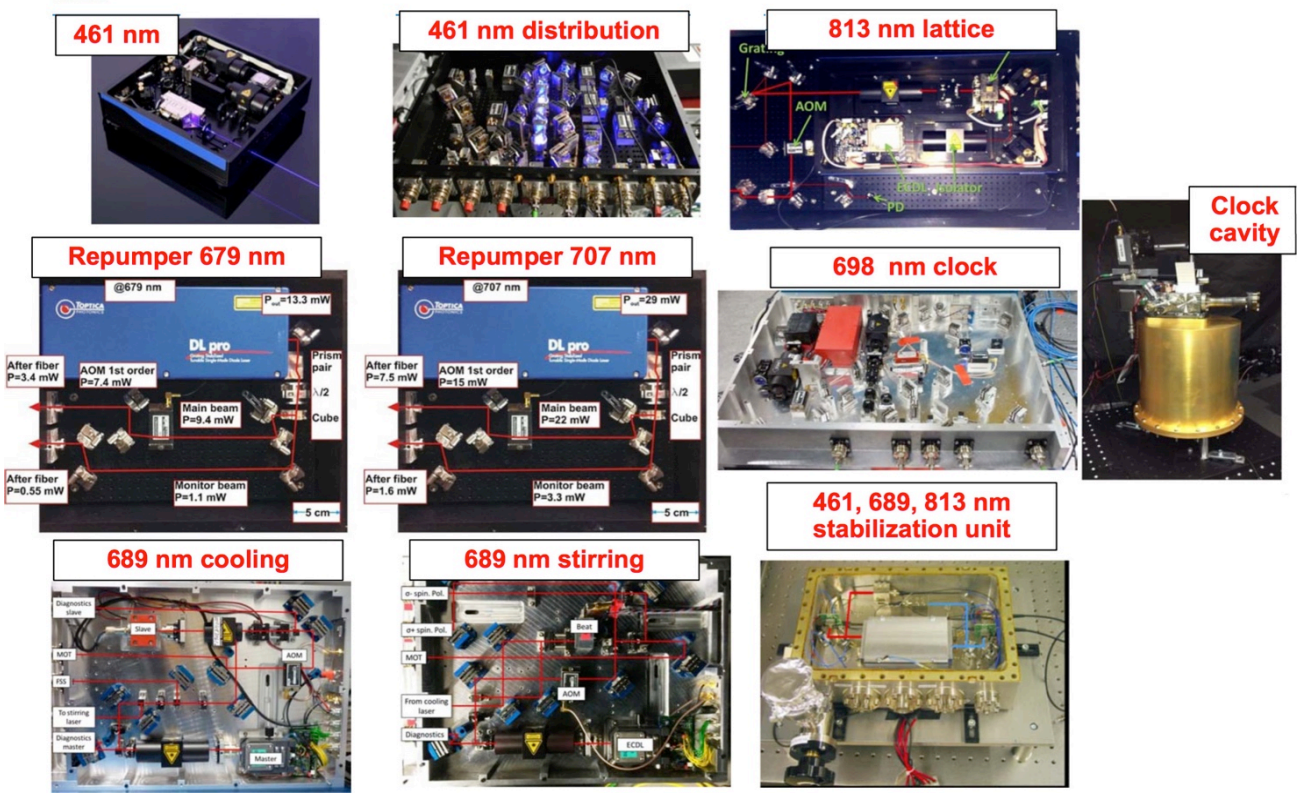

Figure 3. Modular laser system.

## 2.2 Atomics package breadboard

The vacuum system (Figure 4) is conceived to be compact, robust and lightweight. The total volume is 143 liter, and the mass is 50 kg including breadboard. Two ion pumps, one close to the oven with pumping speed 25 liter/s and one just after the science chamber with pumping speed 2 liter/s, maintain the vacuum in the $10^{-10}$ mbar range if not gas load from the oven is present. The atomic source is a low-power oven adapted from the previous development of the SOC consortium[17]. It is connected to the rest of the vacuum system with a CF35 electrical feedthrough flange. A tantalum wire is wound around the reservoir that can be filled with up to 6.4 g strontium (enough for about 10 years of operation). The nozzle of the oven is composed of custom made micro-tubes to collimate the atomic beam: the resulting beam divergence is 20 mrad. Under typical operating conditions with $^{88}$Sr, the power consumption of the oven is about 15 W when the temperature is between 350 and 400°C. An aluminum heat shield around the oven as well as an automated full-closure viewport shutter are used to shield the atoms from the blackbody radiation coming from the oven.

A permanent-magnet Zeeman slower[18] is implemented in the system. The permanent-magnet concept allows having a good capture efficiency without power consumption. The Zeeman slower is 30 cm long and is designed to ensure a rapid decay of the stray magnetic field towards the science chamber.

The science chamber is compact and lightweight. The overall size is $13.0 \times 5.0 \times 2.2$ cm$^3$ and is realized in titanium. Titanium was chosen for its non-magnetic properties and also because its thermal expansion coefficient is similar to that of BK7 glass used for the viewports. Furthermore, titanium is lightweight: the mass of the science chamber is only 208 g. The chamber consists of 8 viewports, 6 small ones on the side used for 2 of the MOT beams (with retroreflection) and for the clock and lattice laser, and two large ones (40 mm), with a total of 10 optical accesses, used for the remaining MOT beams, the repumpers, the beams used for optical pumping and for the detection and imaging (using a PMT and a CCD camera). The viewports are sealed to the chamber using indium.

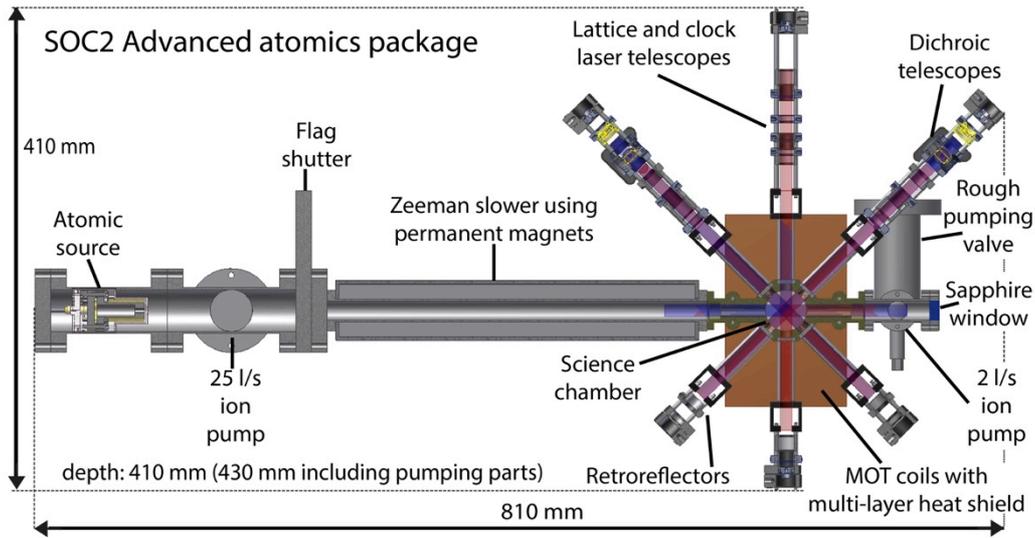

Figure 4. Schematic of the SOC2 atomics package.

With such a compact design for the vacuum chamber, also the coils used to generate the MOT field gradient can be small. The internal and external radii are, respectively, 23 mm and 42 mm. This results in a power consumption of only 6 W for each coil under normal operating conditions, allowing operating the system without need for active cooling. However, for clock operation the temperature of the vacuum system must be controlled. For this purpose a multi-layer heat shield around the science chamber was implemented, consisting of a first copper layer where the coils are mounted, an insulating layer and an internal copper layer close to the vacuum chamber. A thermoelectric cooling (TEC) system can be implemented by placing some TEC elements between the internal and external cooling plate in order to control the temperature of the layer closer to the vacuum system.

### 2.3 Clock laser

The clock laser consists of the laser head, the breadboard and the reference cavity. The laser head (LUH) is a fiber-coupled interference-filter-stabilized External-Cavity Diode Laser (ECDL) with a linewidth of 10 kHz and it is fixed inside the distribution breadboard (size 60 cm × 45 cm × 12 cm, mass 20 kg): four outputs are generated, one going to the cavity for the locking, one to the wavemeter for monitoring the frequency, one to be used for the optical frequency comb and finally one to the atoms. A fiber stabilization is implemented for the fibers going to the atoms and for the one to the frequency comb.

The ultrastable reference cavity is a 10 cm-long vertically mounted ULE cavity. The cavity is enclosed in three gold-coated aluminum shields to reduce the sensitivity to temperature fluctuations. To minimize the influence to vibration the cavity is mounted vertically. The volume of the cavity is 5 liter and the weight is 9 kg. The lowest measured fractional instability was $1.6 \times 10^{-15}$.

## 3. PRELIMINARY RESULTS

The system is fully operational using the isotopic species $^{88}$Sr. It is the choice for initial work, since it is the most abundant strontium isotope and easier to trap and cool, whereas $^{87}$Sr is preferred as optical frequency standard.

### 3.1 Atoms cooling and trapping

Figure 5 shows a detailed scheme of the sequence for the cooling and the clock spectroscopy, as it is implemented in the experiment.

The power in the MOT beams for the first-stage MOT (461 nm) is about 6 mW each (diameter 10 mm); 30 mW are delivered to the slower beam. About $10^7$ atoms can be trapped in less than 300 ms, at a temperature of 2 – 3 mK. The atoms are then cooled by reducing the power of the MOT beams from 6 mW to 40 – 60 µW, while turning off the Zeeman slower beam. In the last step of the cooling sequence, we slightly reduce the magnetic field gradient (from 45 to 35 G/cm). After this step the temperature of the atoms was measured to be between 400 and 450 µK, which is below the Doppler limit (about 700 µK). To explain this, we note that during the loading and the cooling sequence of the first-stage MOT the beams for the second-stage MOT (689 nm, about 3 mW per beam, same diameter as the first-stage beams) are kept on. This means that in this stage, since there is much more power in the 689 nm beams than in the 461 nm beams, the $^1S_0 - {}^3P_1$ transition is dominant compared with the $^1S_0 - {}^1P_1$ transition: therefore this step can be considered as an intermediate step between the first and the second-stage MOT.

Thanks to the high power in the 689 nm beams there is no need for frequency broadening in order to obtain good capture efficiency: about 50% of the atoms are transferred from the first-stage to the second-stage MOT. In the first step of the second-stage MOT the magnetic field is turned down close to zero for a few ms, then it is gradually increased up to 13 G/cm for the last step, while reducing the power in the beams and the detuning to cool the atoms down to few µK. During the cooling sequence about half of the atoms initially trapped in the second-stage MOT are lost. The total duration of the second-stage MOT is 80 ms, including 40 ms for the settling of the trap before the loading of the lattice.

For the detection of the atoms in the second-stage MOT, as well as for the lattice, a resonant (461 nm) beam is used, which is coupled in the same fibers as the MOT beams for the $x$ and $y$ directions. The resonant beam is overlapped with the MOT beams inside the distribution module. In the direction of the horizontal MOT beam ($z$ axis) no resonant beam is used, because this direction is close to the one of the PMT, leading to a large background signal. This configuration for the detection was used in order to reduce the number of optical components mounted around the vacuum system.

The lattice is oriented vertically and has a waist size of 35 µm, with a power of about 170 mW, leading to a trap depth of 150 times the recoil energy. The lattice laser is on all the time, so that after the last step of the second-stage MOT, it is enough to turn off the 689 nm beams to transfer the atoms into the dipole trap. Since in the second-stage MOT the temperature of the atoms is about 2 – 3 µK, good transfer efficiency is achieved (10% or more), so that finally we can trap more than $10^5$ atoms into the lattice. Their lifetime is about 1 second. The lifetime depends strongly on the pressure in the system which, due to the small pumping speed (2 l/s) of the ion pump close to the science chamber, is quite sensitive to the temperature of the oven.

It is remarkable that the atoms can be cooled and trapped into the lattice in less than 400 ms. This allows obtaining high duty cycle, an important factor in the reduction of the Dick effect for the clock instability.

### 3.2 Clock spectroscopy in the optical lattice

The light from the clock laser breadboard is delivered via optical fiber to the vacuum system, where it is overlapped with the lattice laser using a mirror that reflects 99% of the lattice light and transmits about 90% of the clock laser beam. In the future, the reflected component can be used for the fiber noise cancellation or for monitoring the presence of vibrations in the system (i.e. on the lattice) that may induce some Doppler effect in the atoms-light interaction during the spectroscopy. Up to 500 µW can be delivered to the atoms. The size of the waist of the clock beam is about 65 µm, 20 µm larger than the lattice laser waist: this gives a maximum intensity of about 4 W/cm$^2$. Working with $^{88}$Sr, whose clock transition is strongly forbidden, we need to induce it by applying an external bias magnetic field. This bias field is produced by inverting the current in one of the MOT coils, so that a uniform field is generated in the trapping region.

As shown in Figure 5, once the atoms are trapped in the lattice, a waiting time of 30 ms is introduced before turning on the clock laser and the bias field to make sure that the untrapped atoms have flown away. The interrogation time may vary depending on the Rabi frequency (see below): since at the moment we are working with Rabi frequencies of about

10 Hz, an interrogation time of 100 ms is used. After the spectroscopy beam is turning off, the fraction of atoms in the excited state of the clock transition is measured using a shelving detection scheme. First, the atoms in the ground state are detected using a short (10 ms long) pulse of resonant 461 nm light. This process heats up the ground state atoms, which therefore are lost from the trap, leaving only the atoms still in the excited state ("excited-state atoms"). To transfer these atoms to the ground state, the repumpers are turned on for 10 ms. Once in the ground state, the atoms can be detected. The excitation probability originally produced by the clock laser is then calculated as the ratio between the number of detected excited-state atoms and the total number of atoms in the lattice (sum of detected excited-state atoms and detected ground-state atoms).

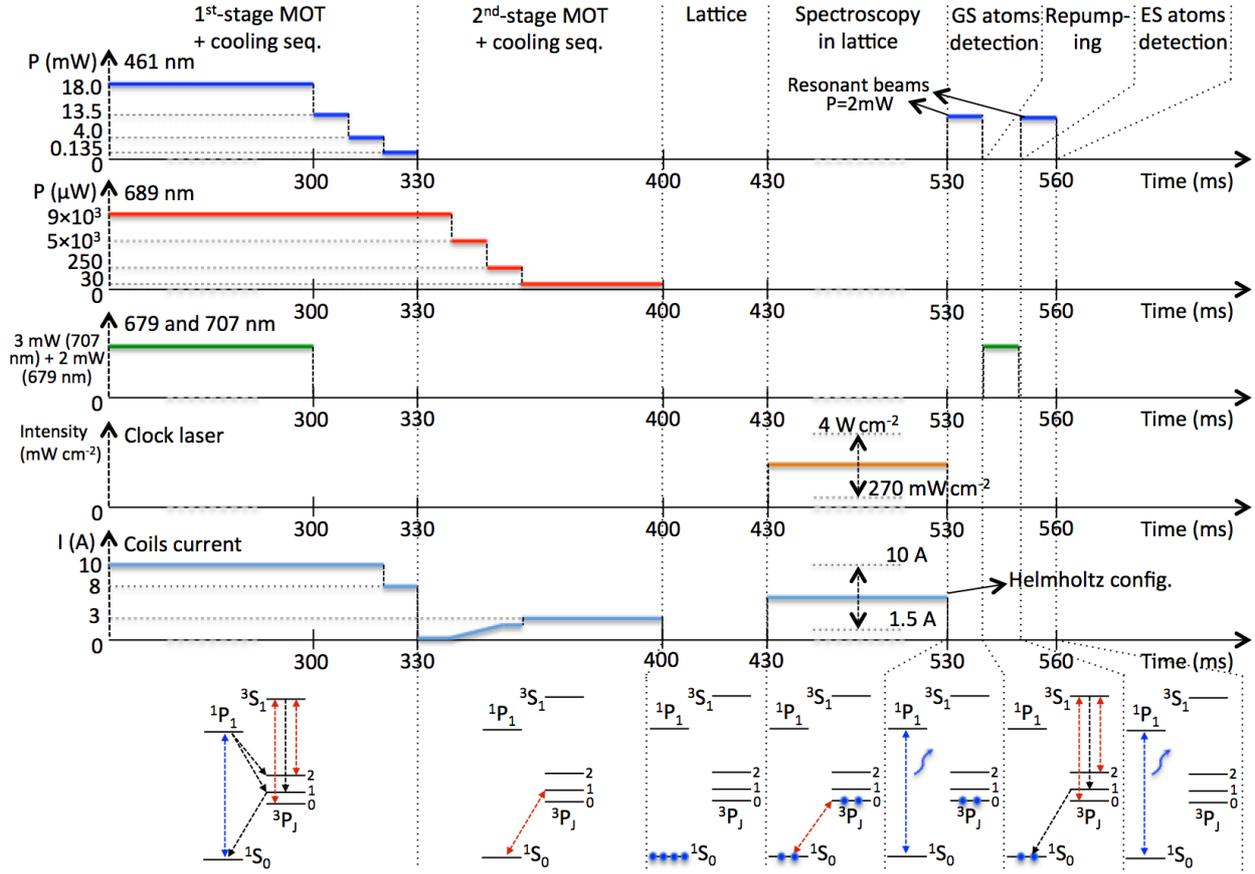

Figure 5. Sequence used for a single cycle of cooling and clock spectroscopy of the Sr atoms. The vertical axes are not to scale. The bottom line shows the atomic transitions occurring in the various stages. The power for the first and second-stage MOT is the total power in the three MOT beams. The lattice laser is not depicted here since it is on during the whole sequence. Regarding the current in the coils, a value of 10 A corresponds to a gradient in the center of the trap of 44 G/cm; for the bias field (Helmholtz configuration) 10 A correspond to 19.6 mT. The intensity of the clock laser as well as the amplitude of the bias field for the clock spectroscopy depends on the application: for high accuracy spectroscopy low intensity and amplitude are used, whereas for the detection of the motional sidebands higher intensities and amplitudes are required. GS: ground state; ES: excited state.

A typical clock transition septcrum is shown in Figure 6. Due to the not ideal vacuum conditions in the SOC2 reference cavity, the measured fractional Allan deviation of the clock laser was approximately $1 \times 10^{-14}$ and this limits the atomic linewidth that can be achieved (left panel). To reduce the laser and thus atomic linewidth, the frequency of the SOC2-cavity-locked laser was additionally locked to a non-transportable, ultrastable reference cavity[22]. The narrower clock laser spectrum then enabled linewidths below 10 Hz (Figure 6, right). The interrogation time is set according to the Rabi frequency (see below) to maximize the excitation probability: the lower the Rabi frequency, the longer the interrogation time. Since the interrogation time is inversely proportional to the Fourirer-limited linewidth, longer probe pulses, and thus smaller Rabi frequencies, are needed to obtain narrow spectral lines. The expected Rabi frequency is $\Omega_R = \alpha \sqrt{I} |B|$, where $\alpha(Sr) = 198 \text{ Hz}/(T\sqrt{\text{mW/cm}^2})$, $I$ is the intensity of the clock laser light and $B$ is the bias field. Therefore, to obtain spectral linewidths in the range of 10 Hz, low bias fields and clock laser intensities are used.

It is worth mentioning that the short-range interactions between the $^{88}$Sr atoms can cause not only a shift in frequency of the transition, but also a broadening[23]. This effect was noticed when the linewidth was below 20 Hz: in order to further reduce the linewidth, it was necessary to reduce the number of atoms in the lattice approximately by a factor of 10 (from $10^5$ to about $10^4$ atoms).

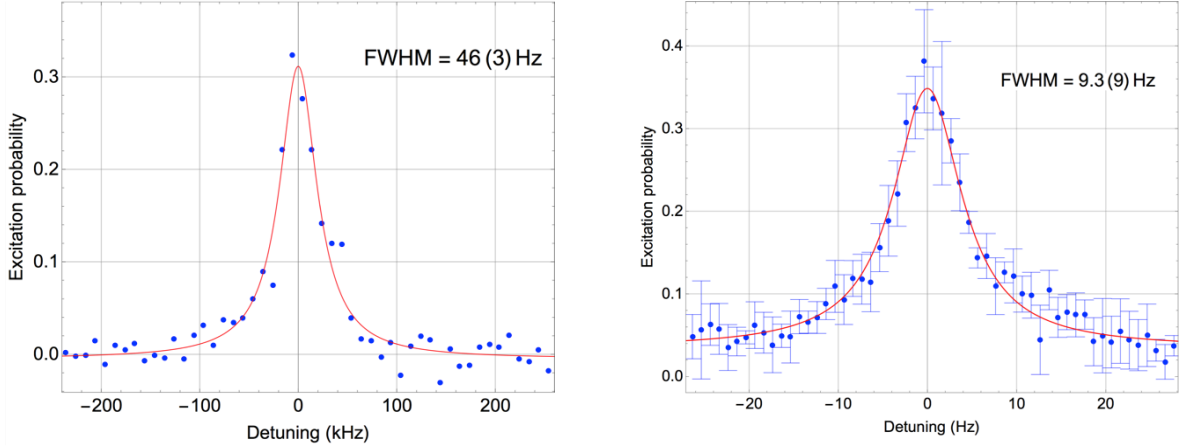

Figure 6. Clock transitions in $^{88}$Sr. Left: using the SOC2 cavity as reference. Clock laser intensity 270 mW/cm², bias field 6 mT, interrogation time 100 ms. Right: by locking the clock laser additionally to an ultrastable stationary cavity. Clock laser intensity 270 mW/cm², bias field 3.9 mT, interrogation time 150 ms.

In the transition shown in Figure 6 the excitation probability is lower than 50%. In order to further investigate this aspect, Rabi oscillations were observed by varying the duration of the probing time (Figure 7). From this data, we obtain two important insights. First, we can measure the Rabi frequency, 33.7(3) Hz. The expected Rabi frequency, with $I = 3 \text{ W/cm}^2$ and $B = 6$ mT, is 65 Hz. The fact that the Rabi frequency obtained from the fit is about half the expected value may be a consequence of an imperfect overlap between the clock laser and the lattice laser beams, so that the clock laser beam is not centered on the center of the atomic distribution. This inconsistency should be removable by checking the overlap of the two beams close to their focus; currently it is only done in the far field. Second, several Rabi oscillations are resolved before they are damped out, which is a sign that the interrogation is performed with a homogeneous excitation probability over the interrogation region.

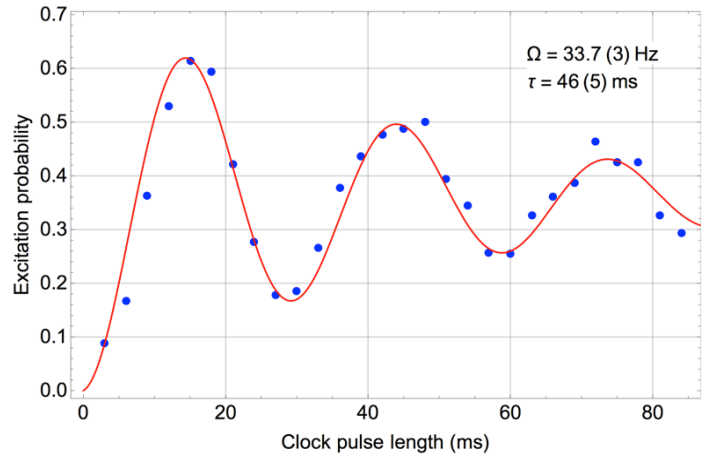

Figure 7. Rabi oscillations of the excited state probability. Clock laser intensity 3 W/cm², bias field 6 mT. The function used for the fit is $a(1 - \cos(2\pi\Omega\Delta t)\exp(-\Delta t/\tau))$, where $\Delta t$ is the clock laser pulse length, $\Omega$ is the Rabi frequency and $\tau$ is the decay time.

Another important source of information is the sideband spectrum (Figure 8). The sharp edge of the higher-frequency (blue) sideband gives an estimation of the longitudinal trap frequency $\nu_z$, here about 53 kHz. This value has to be compared with the expected value coming from the parameters of the lattice beam. With a power of about 170 mW and a waist radius equal to 35 μm, it is 84 kHz. This observed lower value may arise because the atoms are not located in the waist of the lattice (due to a not perfect overlapping between the lattice waist and the MOT beams) or because of some inhomogeneity in the atoms' spatial distribution. Increasing the diameter of the lattice beam may help to have a better control of its overlap with the MOT beams as well as with the clock laser. From the relative amplitude of the red ($a_R$) and blue ($a_B$) sideband one deduces the mean vibrational quantum number $\langle n \rangle$ as $a_B/a_R = (\langle n \rangle + 1)/\langle n \rangle$. Using $\langle n \rangle = \exp(-h\nu_z/(k_B T))/(1 - \exp(-h\nu_z/(k_B T)))$ the temperature $T$ of the atoms can be evaluated. From the data in Figure 8 we obtain $\langle n \rangle = 0.22$ and $T = 1.5\ \mu K$: this low temperature indicates an efficient cooling in the second-stage MOT.

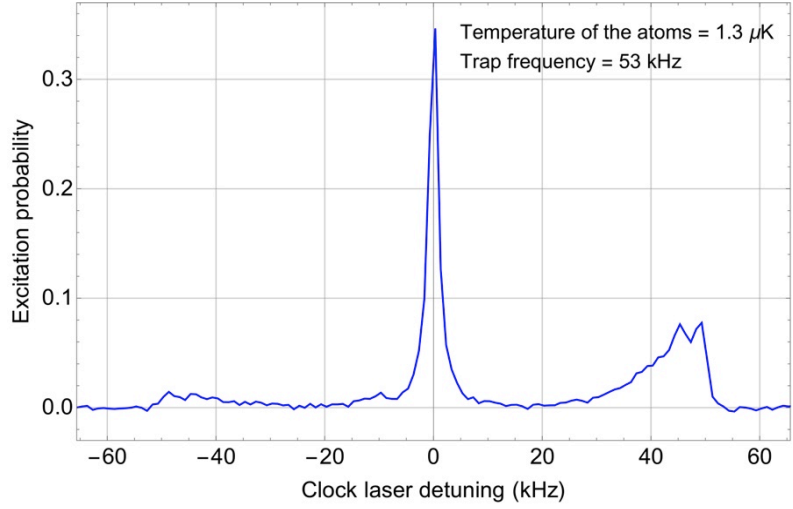

Figure 8. Sideband spectrum. Clock laser intensity 4 W/cm², bias field intensity 19 mT.

## 4. CONCLUSIONS

A compact, robust and lightweight strontium optical lattice clock was developed. It is operating reliably with $^{88}$Sr, showing good cooling and trapping efficiency. More than $10^5$ atoms can be loaded into the lattice with a temperature of 1.3 μK. The spectroscopy of the clock transition was performed showing linewidths as small as 9 Hz. Rabi oscillations as well as sideband spectrum were acquired, showing a good atom coherence time and cooling efficiency.

### 4.1 Outlook

After the formal end of the SOC2 project, the development of the strontium breadboard will go on, with unchanged goals of achieving relative inaccuracy of $5\times10^{-17}$ and instability of $1\times10^{-15}$ at one second interrogation time. To achieve these goals we plan to implement some modification on the current setup.

The first step will be to move to $^{87}$Sr, which should be straightforward once the procedures for the cooling and the spectroscopy of $^{88}$Sr have been optimized. The stirring laser is already available in our laser system. Higher oven temperatures compared with $^{88}$Sr will be needed, and the influence on the atom lifetime in the trap needs to be studied. Second, control of the blackbody radiation shift is necessary. This will be achieved by keeping the temperature of the vacuum system stabilized near room temperature by controlling the temperature of the cooling plates. A TEC system has already been designed and once installed will allow to control the temperature with an inaccuracy below 1.0 K, corresponding to an uncertainty at $7 \times 10^{-17}$ level. Third, to control the electric field in the vacuum chamber, the current windows will be replaced with ITO-coated windows. Fourth, to improve the clock laser stability, the pressure in the clock laser reference cavity needs to be improved, by re-sealing the cavity and increasing the pumping capacity. The goal is to reach the thermal noise level of $5\times10^{-16}$.

To enhance robustness and transportability, the cage system connected to the vacuum chamber (Figure 4) will be made more compact, through the use of fiber-based components to combine the beams. The current breadboards will be exchanged with units using commercial, high-stability mirror mounts, and active temperature stabilization.

An evaluation of the performance of the clock in terms of accuracy and instability will be performed during the year 2016. Once the goal performances have been achieved, the clock will be tested by transport to other locations providing information about the robustness of the apparatus. One concrete application is as a transportable ground clock for the ACES mission. The apparatus will also be used as testbed for testing new subsystems aimed for the space clock and currently under ESA-funded development. These can be rather easily implemented into the existing apparatus thanks to its modularity. Finally, a phase-A study for the ESA mission SOC is envisaged to start in 2016; for this, the SOC Sr breadboard provides an important reference model.

## ACKNOWLEDGEMENTS

The research leading to these results has received initial funding by ESA, DLR and other national sources. Current funding has been provided by the European Union Seventh Framework Programme (FP7/2007–2013) under grant agreement No. 263500. The work at PTB was also funded by the European Metrology Research Programme (EMRP) under IND14. The EMRP is jointly funded by the EMRP participating countries within EURAMET and the European Union. J.H. and D.S. acknowledge the funding from the EPSRC (EP/L001713/1) and Qtea (FP7-People-2012-ITN-Marie-Curie Action "Initial Training Network (ITN)"), respectively. S.O. and S.V. were funded by the Marie-Sklodowska-Curie Action ITN "FACT". S.V. acknowledges funding from the DFG within the RTG 1729. K.B. and Y.S. are members of the UK NQT Hub in Sensors and Metrology and acknowledge funding from EPSRC (within grant EP/M013294/1).